\documentclass[aps,prd,amsmath,amssymb,twocolumn,superscriptaddress,nofootinbib,
showpacs,10pt]
{revtex4}
\usepackage{epsfig,float}
\usepackage{graphicx}
\usepackage{dcolumn}
\usepackage{bm}
\usepackage[usenames]{color}

\newcommand{\y}{Y(4260)}
\newcommand{\juelich}{Institut f\"ur Kernphysik, Institute for Advanced
Simulation and J\"ulich Center for Hadron Physics,\\ Forschungszentrum
J\"ulich, D-52425 J\"ulich, Germany}
\newcommand{\bonn}{Helmholtz-Institut f\"ur Strahlen- und Kernphysik and Bethe
Center for Theoretical Physics,\\ Universit\"at Bonn,  D-53115 Bonn, Germany}
\newcommand{\ihep}{Institute of High Energy Physics and Theoretical Physics
Center for Science Facilities,\\ Chinese Academy of Sciences, Beijing 100049,
China}

\begin{document}

\title{\bm{$Y(4260)$}: hadronic molecule versus hadro-charmonium interpretation}

\author{Qian~Wang}
\affiliation{\juelich}
\author{Martin~Cleven}
\affiliation{\juelich}
\author{Feng-Kun~Guo}
\affiliation{\bonn}
\author{Christoph~Hanhart}
\affiliation{\juelich}
\author{Ulf-G.~Mei{\ss}ner}
\affiliation{\bonn}
\affiliation{\juelich}
\author{Xiao-Gang~Wu}
\affiliation{\ihep}
\affiliation{\juelich}
\author{Qiang~Zhao}
\affiliation{\ihep}

\begin{abstract}
In this paper  we confront both the hadronic molecule and the
hadro-charmonium interpretations of the $\y$ with the experimental data
currently available. We
conclude that the data  support the $\y$ being
dominantly a $D_1\bar D+c.c.$ hadronic molecule while they challenge the
hadro-charmonium interpretation. However, additional data is necessary 
to allow for stronger conclusions.
\end{abstract}
\pacs{14.40.Rt, 14.40.Pq}

\maketitle

In the past decade, a lot of new states, called $X$, $Y$ or $Z$,
were observed in the heavy quarkonium mass region~\cite{Brambilla:2010cs}. Quite
a few of them are close to open-flavor meson-meson thresholds, and can hardly
be accommodated in the traditional quark model.   Among these states,
the charged charmonium-like (or bottomonium-like) states are intriguing
as they are made of at least four quarks. After Belle's observations of
the $Z_b(10610)$ and $Z_b(10650)$ close to $B\bar B^*+c.c.$ and $B^*\bar B^*$
thresholds, respectively~\cite{Belle:2011aa}, the BESIII Collaboration recently
discovered their
possible analogues in the charmonium mass region, the
$Z_c(3900)$~\cite{Ablikim:2013mio},
$Z_c(4025)$~\cite{Ablikim:2013emm} and
$Z_c(4020)$~\cite{Ablikim:2013gca}. The  $Z_c(3900)$ was soon confirmed by the
Belle Collaboration~\cite{Liu:2013dau} and an analysis based on the  CLEO-c
data~\cite{Xiao:2013iha}.

In Ref.~\cite{Wang:2013cya}, it was argued that the strong signal of
$Z_c(3900)$, being a $D\bar D^*+c.c.$ molecular
state~\cite{Wang:2013cya,Guo:2013sya,Wilbring:2013cha}, in the $Y(4260)$ decays
can be explained by a dominant $D_1D$
molecular component in the $Y(4260)$ wave
function~\cite{Ding:2008gr,Li:2013bca}\footnote{Here, $D_1D$ is
a short
notation for $D_1\bar D+c.c.$  For brevity, the same convention will be used
for the $D_1D^*$ and $D_2D^*$ below.}. Related discussions can also be found in
Refs.~\cite{Guo:2013zbw,Liu:2013vfa} emphasizing different aspects. 

Recently,
the interpretation of the $Y(4260)$ as a $D_1D$ molecular state was challenged
in Ref.~\cite{Li:2013yka}, where it is suggested that the $Y(4260)$ is
hadro-charmonium state (a compact quarkonium surounded
by light quarks)~\cite{Voloshin:2007dx,Dubynskiy:2008mq}.
The argument is based on the fact that
 the production of a
pair of $S_L^P=(1/2)^-$ and $S_L^P=(3/2)^+$ heavy mesons, where $S_L$ is
the sum of the spin of the light quark and the orbital angular momentum in
the heavy meson, in electron-positron collisions is forbidden in the heavy
quark limit---in the real world this should translate to a suppressed production
of both the $D_1D$ continuum as well as $D_1D$ molecular states. 

In this paper we confront both interpretations  of the $\y$ the hadro--charmonium
as well as the hadronic molecule
with the data currently available. Especially,
we argue that the $D_1D$ molecular
interpretation of the $Y(4260)$ does not contradict the current
experimental facts despite the suppression of the production of the $D$ and
$\bar D_1$ pair. It is shown that the heavy quark spin symmetry (HQSS) breaking due
to a finite charm quark mass is important in this case~\cite{Li:2013ssa}. We also
discuss the challenges that both interpretations still face.

In Ref.~\cite{Li:2013yka}, Li and Voloshin stressed that in the heavy quark limit the
production of a heavy state in $e^+e^-$--collisions  proceeds via the
electromagnetic current $\bar c \gamma_\mu c$ leading to a
$c\bar{c}$ pair in a $^3S_1$ state with the spin of the heavy system $S_H=1$. At the same time  the 
total angular
momentum of the light degrees of freedom should be $S_L=0$. In the heavy
quark limit both $S_H$ and $S_L$ are conserved.  However, the
light quark total angular momenta in the $S$--wave states $D_1 D^{(*)}$ and $D_2  D^*$,
 $S_L=3/2$ and $S_L=1/2$, can not combine to $S_L=0$.
 Consequently,
 the $S$-wave production
of  $D_1 D^{(*)}$ and $D_2  D^*$ pairs with $J^{PC}=1^{--}$ in 
electron-positron collisions breaks the HQSS and thus would be forbidden, if the
charm quark were infinitely heavy.

 This can also
be understood using the angular momentum decomposition of the heavy and light
degrees of freedom. The $J^{PC}=1^{--}$ states with $(\frac {3}{2})^+$ and $(\frac{1}{2})^-$ components can be
decomposed into the following states~\cite{Voloshin:2012dk,Li:2013yka,Liu:2013rxa}
\begin{eqnarray}\nonumber
\psi_{mn}&\equiv&\frac{1}{\sqrt 2} ([[c\bar c]^{m},[\bar{q}_lq]^{n}]^1-[[\bar c c]^{m},[q_l\bar q]^{n}]^1),
\label{eq:eigenstate}
\end{eqnarray}
where the subscript $l$ denotes the orbital angular momentum carried by  the light quark in the $D_1$ or $\bar D_1$. 
As a result, the $D_1 D$ wave
function projected to the proper quantum numbers reads~\cite{Li:2013yka}
\begin{eqnarray}
|D_1D(^3S_1)\rangle^{1^{--}}=\frac{1}{2}\psi_{01}+\frac{1}{2 \sqrt 2}\psi_{11}+\frac{\sqrt
5}{2\sqrt 2}\psi_{12} \ .
\label{eq:Y(4260)}
\end{eqnarray}
Since $\psi_{10}$ is absent, in the heavy quark limit the $D_1 D$ state
should not couple to the photon.
In addition, in Ref.~\cite{Li:2013yka} it was shown
that neither the rescattering due to the process $D^*\bar D^*\to D_1\bar D$ 
nor the mixing of the
$D_1(2420)$ with the $D_1(2430)$ can evade the above mentioned suppression.
In Ref.~\cite{Li:2013yka} also an attempt was made to quantify a possible
kinematic effect that might increase the amount of HQSS violation
by estimating at the square of the ratio of the $D$-wave amplitude
to the $S$-wave one~\cite{Li:2013yka}
\begin{equation}
  \left| \frac{D}{S} \right|^2 = \frac{(E-2m_c)^2}{2(E+m_c)^2}
  \label{eq:ds}
\end{equation}
which turns out to be about 0.02 at $E=4.26$~GeV although the
energy is already well above $2m_c$. Thus, they conclude that the $S$-wave
production of the $(3/2)^+$ and $(1/2)^-$ charmed meson pairs is heavily
suppressed.

In what follows we will demonstrate that the presence of the suppression
described above does not allow one to exclude that the $Y(4260)$ is a $D_1D$
bound system. On the contrary: all properties of the $Y(4260)$ are
consistent with its molecular interpretation in the presence of a suppressed
production in $e^+e^-$ collisions.

Naively, the HQSS breaking effect is characterized as
$\mathcal{O}(\Lambda_\text{QCD}/m_Q)$, which presents a significant
suppression when $m_Q\gg \Lambda_\text{QCD}$. There is no ambiguity if the heavy
quark mass $m_Q$ is much larger than $\Lambda_\text{QCD}$ so that a finite
numerical factor would not change the suppression much. However, this is not the
case for the charm quark. For the production of the $D_1$ and $\bar D$ pair
around the energy 4.26~GeV, the excess energy $E_e=E-2 m_c$ is not small compared
with the charm quark mass, and might cause a large HQSS breaking.
Equation~\eqref{eq:ds} discussed in Ref.~\cite{Li:2013yka} only represents the $D/S$
ratio for the free charm quark pair. However, HQSS breaking could happen after this in
the nonperturbative hadronization process. For instance, we may think of the
light quark and antiquark pair being produced through bremsstrahlung gluons
radiated from the charm quark. If the gluons carry an energy which is not
negligible compared with the charm quark mass, the spin of the charm quark has a
certain probability to be flipped due to the chromomagnetic interaction $\bar
c\, \vec{B}\cdot \vec{\sigma}\, c/m_c$, where $B^i=\epsilon^{ijk}\partial^jA^k$
and $A^k$ are the gluon fields, and $\vec{\sigma}$ are the Pauli matrices. Thus,
the HQSS breaking amplitude should be proportional to $E_g/m_c$, where $E_g$ is the
gluon energy. The effect in question could reach $E_g/m_c\sim
(M_{D_1}-m_c)/m_c\sim0.6$ in the production amplitude\footnote{In a perturbative
estimate one would need to multiply the estimate with $\alpha_s/\pi$. However,
at the energy just estimated $\alpha_s$ is already large.}, which is numerically
larger than $\Lambda_\text{QCD}/m_c$ by a factor of 2 or 3,  and $\sim
0.3\text{--}0.4$ in the cross section. Therefore, the
suppression in the $S$-wave production of the $D_1 D$ pair in $e^+e^-$
collisions does not need to be as strong as reported in
Ref.~\cite{Li:2013yka}.

In line with the expectation that the HQSS breaking could be sizeable when the
excess energy $E_e=E-2m_c$ is not small in comparison with the charm quark
mass, HQSS is indeed badly broken in many cases for the charmonium states
above the open-charm thresholds. One good example is
the electron decay widths  $\Gamma_{ee}$ of the vector charmonium states. In
the
heavy quark limit, only the $S$-wave heavy quarkonium states are allowed to
decay into an electron-positron pair via a virtual photon. Thus, the value
of $\Gamma_{ee}$ of a $D$-wave heavy quarkonium, which corresponds to the
$\psi_{12}$ state in the decomposition in Eq.~\eqref{eq:eigenstate}, should be
much smaller than that of an $S$-wave state corresponding to $\psi_{10}$. From
Table~\ref{tab:coupling}, one can see that the $\Gamma_{ee}$ value of the
$\psi(3770)$ is one order of magnitude smaller than that of the $\psi(3686)$,
which is consistent with the fact that the $\psi(3770)$ and $\psi(3686)$ are
mainly a $D$-wave state and an $S$-wave state, respectively. However, all the
three states above 4~GeV, among which at least one is $D$-wave state, have
similar $\Gamma_{ee} $. This indicates a strong
HQSS breaking effect in this energy range, which could be caused by either a
mixing between $S$-wave and $D$-wave states (see, e.g.~\cite{Badalian:2008dv})
or due to enhancement resulting from nearby
thresholds~\cite{Voloshin:2012dk}.~\footnote{In principle, these two different
scenarios are not independent because coupling to the nearby open-charm channels
can induce mixing. } The mixing angle between the $2D$ and $3S$ states could be as
large as 34$^\circ$ in Ref.~\cite{Badalian:2008dv}.

Another observation supporting that spin symmetry violations
are potentially significant for the $\y$ is that
$|(M_{D_1}+M_D)-M_Y|\sim 30\,$MeV,
$|(M_{D_1}+M_{D^*})-M_Y|\sim 160\,$MeV and $|(M_{D_2}+M_{D^*})-M_Y|\sim 200\,$MeV---
in the heavy quark limit all mentioned thresholds were degenerate.

\begin{table}[t!]
\caption{\label{tab:coupling}Electron widths of the
vector charmonium states~\cite{Beringer:1900zz}. }
\begin{ruledtabular}
\begin{tabular}{lc}
 Resonance & $\Gamma_{ee}~(\mathrm{keV})$\\
 \hline
 $J/\psi$ & $5.55\pm 0.14$  \\
$ \psi(3686)$ & $2.35\pm0.04$ \\
 $\psi(3770)$ & $0.262\pm 0.018$  \\
 $\psi(4040)$ & $0.86\pm 0.07$ \\
 $\psi(4160)$ & $0.83\pm 0.07$\\
 $\psi(4415)$& $0.58\pm 0.07$ \\
 \end{tabular}
 \end{ruledtabular}
 \end{table}

Having established that in the 4~GeV region one should expect a significant
violation of heavy spin symmetry and thus allowing for the production of the
$\y$ in $e^+e^-$ collisions even if it is a $D_1D$ molecule, we now
collect arguments that the data currently available are actually in favor of
this molecular interpretation while they pose a challenge to the hadro-charmonium
picture.

\begin{enumerate}
\item The $\y$ decays into both the $\pi^+\pi^- J/\psi$ and $\pi^+\pi^- h_c$, and
the cross sections for the processes $e^+e^-\to
\pi^+\pi^- h_c$ and $e^+e^-\to \pi^+\pi^- J/\psi$ at 4.26~GeV are
similar~\cite{Ablikim:2013gca}, which implies a large HQSS breaking. Especially,
the data show that a large part of the cross section for the $\pi^+\pi^-h_c$ is
not from the $Z_c$ states.
This is natural in the $D_1D$ molecular picture, as both the $J/\psi$ and $h_c$
couple to the charmed and anticharmed meson pair. Furthermore, once the $D_1D$
pair is produced, the decay into the  $\pi^+\pi^-h_c$
will be facilitated by the charmed meson loops. This is because both the
$\y D_1\bar D$ and the $h_cD^*\bar D$ vertices are $S$-wave, and in this case
there is a large enhancement factor $1/v^3$ with $v\ll1$ being the velocity of
the intermediate charmed mesons~(for detailed discussions, we refer to
Refs.~\cite{Guo:2010ak,Cleven:2011gp,Guo:2013zbw}). In the hadro-charmonium
picture, Li and Voloshin explain the decays of the $\y$ into both the
$\pi^+\pi^- J/\psi$ and $\pi^+\pi^- h_c$ channels by mixing two
hadro-charmonium states whose cores are a $^1P_1$ and $^3S_1$
charmonium, respectively, into the $\y$ and $Y(4360)$~\cite{Li:2013ssa}. However,
the former and the latter state were observed in the $J/\psi\pi^+\pi^-$ 
and in the $\psi'\pi^+\pi^-$ channel, respectively---although
current data do not fully exclude that both states are seen in both transitions
due to limited statistics. 
Because the non-relativistic
wave functions of the $J/\psi$ and $\psi'$ do not overlap, the hadro-charmonium
interpretation faces the difficulty to explain why the $\y$ and $Y(4360)$
are not observed in the same final states. 

Also in the molecular picture it would be natural if the $Y(4360)$ and the $\y$
were close relatives---after all the distance of $\y$ to the $D_1 D$ threshold
is very similar to the distance of $Y(4360)$ to the $D_1D^*$ threshold and $D$ 
\& $D^*$ are spin partners. It remains to be seen if the rates discussed above 
are consistent with such a picture or not. To address this issue requires a microscopic calculation
that we are currently working on.

\item There is a dip in the measured $R$ values around the mass of the $\y$. As
is most clearly demonstrated in Ref.~\cite{Pakhlova}, the dip in this region
emerges after summing up the two-body channels $D^{(*)}\bar D^{(*)}$ and the
three-body channels $D\bar D^{(*)}\pi$ --- none of the individual cross sections
shows a prominent dip. In the hadronic molecular model where the main component
of the $\y$ is $D_1D$, we expect that the $\y$ decays mainly into the $D\bar
D^*\pi+c.c.$ through the decays of the $D_1$ meson. Therefore, the dip behavior
that emerges in the inclusive cross section from the rise of the $DD^*\pi$ rates
beyond 4.2~GeV is fully in line with the hadronic molecular scenario since the
$Y(4260)$ acts as a doorway for the three body final states.
Furthermore, the model does not contradict the upper limit
$\mathcal{B}(\y\to D^0D^{*-}\pi^+)/ \mathcal{B}(\y\to\pi^+\pi^-J/\psi)<
9$~\cite{Pakhlova:2009jv}. On the other hand, in the hadro-charmonium picture
there is no suppression for $e^+e^-\to \y$, and the $\y$ decay dominantly into a
charmonium and two pions. Therefore, one would expect a pronounced peak in the
$R$-ratio around the mass of the $\y$. Especially in this case the $R$ values should
not be saturated by the open charm channels at this energy.  Further
measurements in the $D\bar D^*\pi$ channel would be very helpful to distinguish
between the two models since in the hadro-charmonium model such decays are not
expected to be important.

\item From the discussion above, we expect $\Gamma_{ee}$ of $Y(4260)$ to be in the range of $1\%\sim 10\%$
of $\Gamma_{ee}(J/\psi)=5.55~\mathrm{KeV}$.  Unfortunately, the electron width of the $\y$ cannot be calculated
model-independently. We may estimate it by assuming that the photon coupling to
the $\psi_{12}$ component of the $D_1D$ molecule to be the same as that to the
$2\,^3D_1$ charmonium state which is the closest $D$-wave state.
If we take $\Gamma_{ee}(2^3D_1)=0.059$~keV calculated in
Ref.~\cite{Badalian:2008dv}, taking into account the decomposition given in
Eq.~\eqref{eq:Y(4260)}, we have
$  \Gamma_{ee}(Y(4260)) \simeq \frac58 \Gamma_{ee}(2^3D_1) = 37~\text{eV}$,
which is much smaller than the existing upper limit 580~eV~\cite{Mo:2006ss} about ten percent of the $e^+e^-$ decay width of $J/\psi$. Alternatively, we can extract the upper limit of $\Gamma_{ee}(Y(4260))$ from the formula
$\sigma_{e^+e^-\to D^0D^{*-}\pi^+}=12\pi\Gamma_{ee}(Y(4260))\Gamma_{Y(4260)\to D^0D^{*-}\pi^+}|G_Y(s)|^2$ with $G_Y(s)$~\cite{Cleven:2013mka} the propagator of $Y(4260)$.
Using the experimental cross section $0.1\sim 0.8~\mathrm{nb}$ for $e^+e^-\to Y(4260)\to D^0D^{*-}\pi^+$ from Belle~\cite{Pakhlova:2009jv} and $\Gamma_{Y(4260)\to D^0D^{*-}\pi^+}=\Gamma_{Y(4260)}/6 \simeq 18~\mathrm{MeV}$~\cite{Beringer:1900zz}, we estimate $\Gamma_{ee}(Y(4260))\simeq 100\sim 800~\mathrm{eV}$, which again is within the range of values estimated above.

\item In a molecular picture for the $\y$ it appears natural that the
$Z_c(3900)$ is observed in the decay~\cite{Wang:2013cya}, in line with
observations. On the other hand, in the hadro-chamonium scenario, where the $\y$
is predominantly  a compact charmonium state surrounded by an isoscalar pion
cloud, it appears difficult to understand why the decay into its building
blocks should run via an isovector intermediate state. Analogously, within the
molecular picture for the $\y$ in Ref.~\cite{Guo:2013zbw} it was predicted that
the $X(3872)$ should be produced in $\y$ radiative decays. Also this transition
would be difficult to explain within the hadro-charmonium interpretation. It
should be mentioned that there are preliminary data available from BESIII where
the observation of $\y\to \gamma X(3872)$ was reported~\cite{Ablikim:2013dyn}.

\item In the hadronic molecule picture one expects that the decay chain $\y \to D\bar D_1\to D\bar D^*\pi$  is very important. This would  lead to a prominent peak at the upper end of the $D^*\pi$ invariant mass spectrum due to the intermediate $D_1$~\cite{Cleven:2013mka}. In contrast, one would not expect the same feature in the hadro-charmonium picture.

\item The data  for $e^+e^-\to D^*\bar D^*\pi$~\cite{Ablikim:2013emm}, which is only a factor
of 2-3 below those for $e^+e^-\to D^*\bar D\pi$ at 4260 MeV, provide a challenge to both pictures
for the $\y$
hadro-charmonium as well as molecule. Analogously to the remark at the end of item 1 above,
a microscopic calculation of  $(D_1,D_2)$ scattering off  $(D,D^*)$ is necessary in order
to see if the molecular ansatz is consistent with the data.

\end{enumerate}

\medskip

In summary, although the $( 3/2 )^+ + ( 1/2
)^-$ charmed meson pair production vanishes in the heavy quark
limit~\cite{Li:2013yka}, we claim that the resulting suppression
for the physical charm quark mass
 is not in conflict with the interpretation that the main
component of the $Y(4260)$ is a $D_1D$ molecule. The HQSS breaking effects at
above 4~GeV can be large. We have examined known experimental constraints on the
$\y$, and found that the hadronic molecular model does not contradict these
constraints. On the other hand, we argue that phenomenology challanges
the hadro-charmonium interpretation. Further high luminosity measurement at
BESIII will help us to gain more insights into the nature of the $\y$ and to
strenghten the statements given above stronger.
Especially, a measurement in the $D\bar D^*\pi$ channel with improved
statistics  will help to distinguish
the hadronic molecular model from the hadro-charmonium one.

\begin{acknowledgments}
We are grateful to Misha Voloshin for valuable discussions. This work is
supported in part by the DFG and the NSFC through funds provided to the
Sino-German CRC 110 ``Symmetries and the Emergence of Structure in QCD'', the EU
I3HP ``Study of Strongly Interacting Matter'' under the Seventh Framework
Program of the EU, the NSFC (Grant No. 11165005 and 11035006), and the
Ministry of Science and Technology of China (2009CB825200).
\end{acknowledgments}

\end{document}